\documentclass{revtex4}
\usepackage{amsmath}
\usepackage{amsfonts}
\usepackage{amssymb}
\usepackage{graphicx}
\begin{document}
\title{Neutrino mass determination from a four-zero texture mass matrix.}
\author{J. Barranco}
\email{jbarranc@fisica.ugto.mx}
\affiliation{Division de Ciencias e Ingenier\'ias,  Universidad de Guanajuato, Campus Leon,
 C.P. 37150, Le\'on, Guanajuato, M\'exico.}
\author{D. Delepine}
\email{delepine@fisica.ugto.mx}
\affiliation{Division de Ciencias e Ingenier\'ias,  Universidad de Guanajuato, Campus Leon,
 C.P. 37150, Le\'on, Guanajuato, M\'exico.}
\author{L. Lopez-Lozano}
\email{lao-tse@fisica.ugto.mx}
\affiliation{Division de Ciencias e Ingenier\'ias,  Universidad de Guanajuato, Campus Leon,
 C.P. 37150, Le\'on, Guanajuato, M\'exico.}
\date{\today}
\begin{abstract}
We analyze the different  parametrizations of a general four-zero texture mass matrices for quarks and leptons, that are able to
reproduce the CKM and PMNS mixing matrices. This study is done through a $\chi^2$ analysis. In quark sector, only four
solutions are found to be compatible with CKM mixing matrix. In leptonic sector, using the last experimental results about
the mixing angles in the neutrino sector, our $ \chi^2$ analysis shows a preferred value for $m_{\nu_3}$ to be around
$0.05$ $eV$ independently of the parametrization of the four-zero texture mass matrices chosen for the charged leptons
and neutrinos.
\end{abstract}
\maketitle
\section{Introduction}
The Yukawa sector of the Standard Model (SM)
parametrizes the main phenomenological characteristics of fermions with respect to mass and flavor.
  Although there is increasing information on the numerical values of these parameters, a
fundamental understanding of their origin is currently lacking.
The total number of families, the smallness of neutrino masses,the origin of CP violation, the mass spectrum,
are among the problems that still  have no answer within the SM. A phenomenological approach to these questions  is to assume some general textures for the Yukawa matrices for quarks and leptons   and then to compare with experimental data on flavor mixing. The interesting point in this approach is that the hidden symmetry existing in the proposed textures could give some hints on how to extend the Standard Model gauge symmetries. For each chosen textures for Yukawa matrices, it is possible to relate mixing angles and fermion masses \cite{Joglekar:1976pe,Wilczek:1977uh,Fritzsch:1977za,Leurer:1992wg,Ramond:1993kv,Ibanez:1994ig} and then to compare to our experimental knowledge on fermion masses and mixing \cite{Fritzsch:1977vd,Fritzsch:1979zq,Fritzsch:1999ee,Pakvasa:1977in,Harari:1978yi,Fritzsch:1995nx,Fritzsch:2002ga,Randhawa:2006sq,Dev:2008cj,Gupta:2011zzg}. Most of the studies have been done for the quarks sector but some analysis have been made using a six-zero texture related with neutrinos \cite{Xing:2002sb,Branco:2007nn} using an hermitian mass matrix. These models have been ruled out by experimental bounds on mixing angles \cite{Du:1992iy,Fritzsch:2002ga,Frampton:2002yf,Aaij:2011jh}.
Other kind of textures have been proposed for leptonic sector as the one-zero texture \cite{Merle:2006du,Lashin:2011dn} and single or
double vanishing minors \cite{Lashin:2007dm,Lashin:2009yd,Lashin:2010jb} with relative success.

Another promising texture candidate is the four-zero texture which seems to be a good candidate to reproduce the CKM matrix mixing elements \cite{Fritzsch:1977vd,Fritzsch:1979zq,Fritzsch:1999ee,Fritzsch:2002ga}. Usually, it has been assumed for the four-zero textures that fermion masses matrices are hermitian \cite{Fritzsch:1977vd,Fritzsch:1979zq,Fritzsch:1999ee,Branco:1999nb,Zhou:2003ji,Fritzsch:2002ga,Matsuda:2006xa}. It has been pointed out that all fermions sectors can be described with the same formalism, like in $SO(10)$ Great Unification models \cite{Dev:2012xn,Buchmuller:2001dc,Bando:2003ei} or assuming discrete
symmetry  groups \cite{Pakvasa:1977in,Harari:1978yi,Fritzsch:1999ee,Xing:2003zd}. Hence,  it is natural to consider that the same
texture-mass matrix could be used for both quarks and leptons \cite{Xing:2002ta,Hu:2011ac,Barranco:2010we,Canales:2011ug}. Previous
works have analysed the relation between the quark masses hierarchy and the CKM matrix in hermitian case
for the four-zero texture \cite{Branco:1999nb,Zhou:2003ji,Fritzsch:2002ga} and more recently some studies have included a unified four-zero texture for both the leptonic and the
quark sectors \cite{Matsuda:2006xa,Barranco:2010we,GonzalezCanales:2011zz}.
The later analysis shown that hermitian four-zero texture are compatibles with experimental results on quark and leptonic masses as well as CKM and PMNS mixing angles.

In this work, we shall focus on a general four-zero texture parametrization for quarks and leptons Yukawa matrices without the hermiticity assumption usually done for these matrices. From these general four-zero texture parametrization, we shall extensively study all the solutions obtained for the fermion masses and mixing.  We shall demonstrate that using as input parameters the fermion masses, all the solutions can be described through an extra free parameters for each Yukawa matrices. Then these free parameters will be fine-tuned in order to reproduce  both the Cabibbo-Kobayashi-Maskawa (CKM) and the
Pontecorvo Maki Nakagawa Sakata (PMNS) matrices. The fit is done by a $\chi^2$ analysis and we find that from all the parametrizations the current
data on CKM mixing angles excludes all except four parametrizations. In the leptonic sector, the absolute neutrino masses are unknown. We shall show that in order to
reproduce the PMNS mixing angles the neutrino mass of the heaviest neutrino should be around $\sim 0.05$ eV,
a result that is independent of the parametrization used. From this, we can conclude that in order to
have a good fit on the leptonic mixing angle by assuming a four zero texture
mass matrix the absolute neutrino masses will be fixed.
Our analysis is new compared to previous similar analysis \cite{ Matsuda:2006xa} in three aspects:
\begin{itemize}
\item We update the latest results on leptonic mixing angle $\sin \theta_{13}$ and we show that with this value of $\sin \theta_{13}$, the assumptions to have a four-zero textures for the Yukawa matrices fix the neutrino mass scale to be around 0.05 eV.
\item We  study all possible parametrizations of the mixing angles in terms of the fermion masses.
\item We  avoid any approximations since the precision data on CKM mixing and the PMNS mixing impose a precise fine-tuning of the
free parameters.
\end{itemize}

In addition,  we shall show that the assumption of hermitian four-zero texture matrices is not necessary.
 Actually, within the SM, Yukawa couplings need not be either symmetric or hermitian, therefore it is
interesting to complement these kind of analyses by looking Yukawa couplings that are not
necessarily hermitian \cite{Branco:1994jx}. Furthermore, non-hermitian mass matrices can be obtained
in some models with discrete symmetries, such as the $S_3$ models \cite{Mondragon:2011zz}.

This paper is divided in following sections. In Section \ref{sectionII}, we explicitly show
the process to diagonalize  the general four-zero texture
mass matrices and obtain all possible different parametrizations in terms of the fermion masses.
Then, the mixing matrices are
obtained. In section \ref{sectionIII}, we analyze the different solutions obtained from diagonalization
through a $\chi^2$ analysis
using the up-to-date experimental values for the CKM and PNMS mixing matrices. Finally in section
\ref{conclusions}, we present our conclusions.


\section{Parametrizations of four-zero texture mass matrices}\label{sectionII}

We assume that the Yukawa matrices have a non-hermitian four-zero texture in the flavor basis which is given by
\begin{equation}\label{4zerotexture}
M^f=\left(
\begin{array}{ccc}
 0 & C_f & 0 \\
 C'_{f} & D_f& B_f \\
 0 & B'_{f} & A_f
\end{array}
\right).
\end{equation}
In hermitian case, one has to assume that  $C'_f=C_f^*$ and $B'_f=B_f^*$.  Here we only assume that $|C_f|=|C'_f|$ and $|B_f|=|B'_f|$ but the phases can be different \cite{Branco:1994jx,Mondragon:2011zz}.

The most general diagonalization of Yukawa matrix must be done trough a bilinear transformation. In the case of
three generations:
\begin{equation}
U_L^{f\dagger}M^fU_R^f=\text{diag}(m_1^f,m_2^f,m_3^f),
\end{equation}
where $f=u,d,\ell$ is a flavor index and $\{U^f_L, U_R^f\}$ are $SU(3)$ matrices. By definition, the eigenvalues of
$M^f$ must be the masses of fermions, thus they are real. For a general matrix, the unitary matrices
are found solving the equations
\begin{eqnarray}
U_L^{f\dagger}M^fM^{f\dagger}U_L^f&=&\text{diag}(m_1^{f2},m_2^{f2},m_3^{f2})\label{generaldiagonalization1},\\
U_R^{f\dagger}M^{f\dagger}M^{f}U_R^f&=&\text{diag}(m_1^{f2},m_2^{f2},m_3^{f2})\label{generaldiagonalization2},
\end{eqnarray}
for $U_L$ and $U_R$.
 From now on, we  omit the index $f$ for short.
The $H \equiv MM^{\dagger}$ matrix  can be easily computed with the matrix (\ref{4zerotexture}) as
\begin{equation}
H=\left( \begin{array}{ccc}
|C|^2&CD^*&CB'^*\\
DC^*&|B|^2+|C'|^2+|D|^2&B'^*D+A^*B\\
B'C^*&B'D^*+AB^*&|A|^2+|B'|^2\\
\end{array}\right)\label{matrizH}.
\end{equation}
This is a hermitian matrix and can be diagonalized by an unitary transformation.
Writing $B(B')=be^{i\phi_B(\phi_{B'})}$, $C(C')=ce^{i\phi_C(\phi_{C'})}$, $D=de^{i\phi_D}$ and $A=ae^{i\phi_A}$
it is possible to separate the phases of the non-diagonal elements, throughout the unitary transformation
\begin{equation}
H=P^\dagger\tilde{H}P,
\end{equation}
where $P=e^{-\frac{i}{2}\Xi}\text{diag}\left(e^{\frac{i}{2}\Xi},e^{i(\phi_C-\phi_D)},e^{i(\phi_C+\phi_{B'}+\Xi)}\right)$ and
\begin{equation}
\Xi=\arctan\left[\frac{a\sin(\phi_B+\phi_{B'}-\phi_A-\phi_D)}{d+a\cos(\phi_B+\phi_{B'}-\phi_A-\phi_D)}\right].
\end{equation}
Then we have that $\tilde{H}$ is real and symmetric that depends on four positive parameters and one combination of phases, that is
\begin{equation}
\tilde{H}=\left( \begin{array}{ccc}
c^2&cd&bc\\
cd&c^2+d^2+b^2&b|d+a\delta|\\
bc&b|d+a\delta^*|&a^2+b^2\\
\end{array}\right)\label{matrizHtilde},
\end{equation}
where $\delta=e^{i(\phi_D-\phi_{B'}-\phi_B+\phi_A)}$. The matrix (\ref{matrizHtilde}) can be diagonalized by an orthogonal matrix $\mathcal{O}_{ij}=v_i(m_j^2)$, formed with the $i$ component of the eigenvectors $\mathbf{v}(m_i^2)$ that arises from the solution of $(\tilde{H}-m^2_i1_{3\times 3})\textbf{v}(m_i^2)=\textbf{0}$, where $m_i^2$ are the eigenvalues of $\tilde{H}$. Therefore the unitary matrix that diagonalizes (\ref{matrizH}) is given by $U_L=\mathcal{O}P$. Because the diagonalization of $\tilde{H}$ is performed by an orthogonal transformation, the invariants under this transformations give the system
 \begin{eqnarray}\label{invariants}
 \text{Tr}(\tilde{H})&=&m_1^2+m_2^2+m_3^2, \nonumber\\
 \text{Tr}^2(\tilde{H})-\text{Tr}(\tilde{H^2})&=&2m_1^2m_2^2+2m_1^2m_3^2+2m_2^2m_3^2,\nonumber\\
  \text{Det}(\tilde{H})&=&m^2_1m^2_2m^2_3,
 \end{eqnarray}
that reduce the number of free parameters when we solve for $b$, $c$ and $d$ in terms of the eigenvalues $m_i^ 2$,
the parameter $a$ and the phase $\delta$. Thus the components of the \emph{i}-eigenvector of $\tilde{H}$ is given by
\begin{eqnarray}\label{eigenvectors}
v_1(m_i^2,a,\delta)&=&(b^2+c^2-cd+d^2-m_i^2)(a^2-b|d+a\delta |+b^2-m_i^2)\nonumber\\
&&-b(c-|d+a\delta |)(-b|d+a\delta |+b^2+c^2+d^2-m_i^2), \nonumber\\
v_2(m_i^2,a,\delta)&=&-m_i^2(a^2+b^2+c^2-cd)+c(a^2+bd)(c-d)\nonumber\\
&&+b\left[c(b-c)+m_1^2\right]|d+a\delta |+m_i^4,\nonumber \\
v_3(m_i^2,a,\delta)&=&\left[c(d-c)-m_i^2\right](b^2+c^2+d^2-b|d+a\delta |-m_i^2)\nonumber\\
&&+c(b-d)(b^2+c^2-cd+d^2-m_i^2).
\end{eqnarray}

From a general analysis of the solutions of previous equations, the only solutions corresponding to eigenvalues
given by $m_{1,2,3}^2$, i.e. , independent of the other free parameters, are given only when
$\Xi_f=0$ for $f=u,d,\ell,\nu$.
Using this property, the diagonal matrix of phases can be written as $P^f=\text{diag}(1,e^{i\phi^f_1},e^{i\phi^f_2})$, where $\phi^f_1\equiv\phi_{C_f}-\phi_{D_f}$ and $\phi^f_2\equiv\phi_{C_f}+\phi_{B_f'}$ for $f=u,d,\ell,\nu$. As expected, the hermitian case can be reached if the phases are constrained to $\phi_{D_f}=0$ and $\phi_{B_f}=-\phi_{B'_f}$. It is important to stress that the relations between the elements of the mixing matrix and the masses are not the same than in the hermitian case.

The system (\ref{invariants}) has formally 32 solutions that are closely related by chiral transformations that basically change the sign of the masses $m_i$, thus without loss of generality we can assumed that $m_i>0$, where $i=1,2,3$, and adjust the interval of possible values that the free parameters can take. Once the restrictions on parameters and the normal order of masses $m_1<m_2<m_3$ are introduced, the number of independent solution is reduce to three. This means that exist three independent parametrizations for $U^f_L$ that diagonalize $H$. Such parametrizations  are given by
\begin{equation}\label{par3}
{\rm Parametrization \quad 1}\qquad \tilde{m}_1\leq a'\leq \tilde{m}_2\quad\left\{\begin{array}{ccc}
b'&=&\sqrt{\frac{(a'-\tilde{m}_1)(-a'+\tilde{m}_2)(\tilde{m}_3+a')}{a'}}\\
d'&=&-a'+\tilde{m}_1+\tilde{m}_2-\tilde{m}_3
\end{array}\right.
\end{equation}
\begin{equation}\label{par5}
{\rm Parametrization \quad 2} \qquad \tilde{m}_1\leq a'\leq \tilde{m}_3\quad\left\{\begin{array}{ccc}
b'&=&\sqrt{\frac{(a'-\tilde{m}_1)(a'+\tilde{m}_2)(\tilde{m}_3-a')}{a'}}\\
d'&=&-a'+\tilde{m}_1-\tilde{m}_2+\tilde{m}_3
\end{array}\right.
\end{equation}
\begin{equation}\label{par6}
{\rm Parametrization \quad 3} \qquad \tilde{m}_2\leq a'\leq \tilde{m}_3\quad\left\{\begin{array}{ccc}
b'&=&\sqrt{\frac{(a'+\tilde{m}_1)(a'-\tilde{m}_2)(\tilde{m}_3-a')}{a'}}\\
d'&=&-a'-\tilde{m}_1+\tilde{m}_2+\tilde{m}_3
\end{array}\right.
\end{equation}

where for all cases $c'=\sqrt{\frac{\tilde{m}_1\tilde{m}_2\tilde{m}_3}{a'}}$. Here  the parameters $a'=\frac{a}{m_3}$, $b'=\frac{b}{m_3}$, $c'=\frac{c}{m_3}$ and $d'=\frac{d}{m_3}$ have been defined. Likewise the scaled masses are $\tilde{m}_i=\frac{m_i}{m_3}$ for $i=1,2,3$; here one can see that $\tilde{m}_3=1$.

As seen in the parametrizations (\ref{par3}-\ref{par6}) the factor $a'$ ranges between a minimal $\tilde{m}_{\text{min}}$ and maximal $\tilde{m}_{\text{max}}$ value due to the restriction $b'>0$. This allows to write a linear dependence of $a'$ in terms of $\tilde{m}_i$'s and a free parameter $x_f$:
\begin{equation}
a'_1(x_f)=\tilde{m}_{max}\left(1-x_f\frac{\tilde{m}_{max}-\tilde{m}_{min}}{\tilde{m}_{max}}\right).
\end{equation}
With this we have a complete description for parameters of $M^f$ in terms of $\tilde{m}_i^f$ and $x_f$ for $\delta=1$. The fact that we have three parametrizations for each fermion  mass matrix  (up and down) say that there are 9 possibilities to construct the $V_{\rm CKM}$ and $U_{\rm MNS}$ matrices.

\section{Fitting $V_{\rm CKM}$ and $U_{\rm MNS}$}\label{sectionIII}
\begin{figure}[t]
\includegraphics[width=0.6\textwidth]{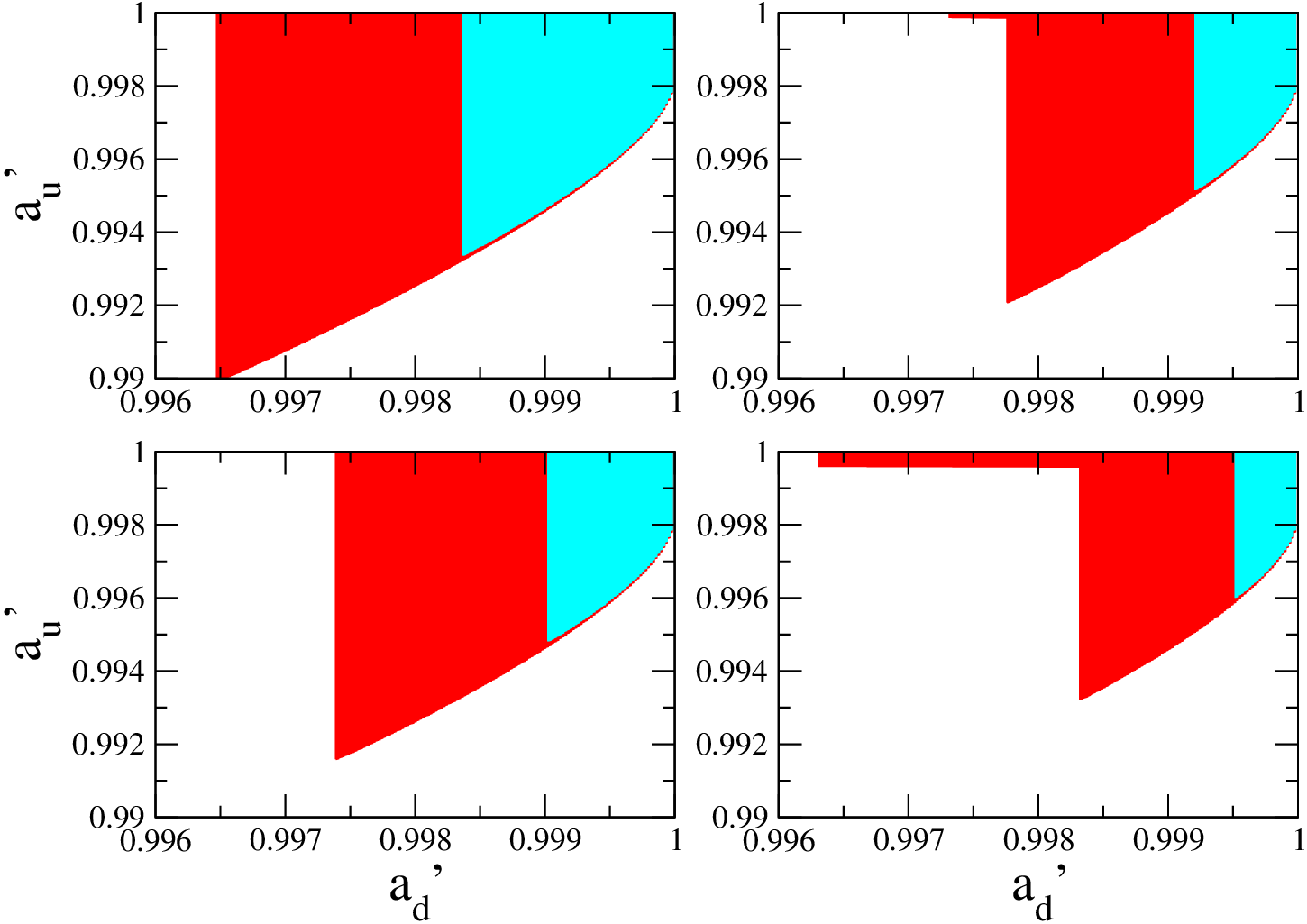}
\caption{Allowed values of parameters $a'_u,a'_d$ that fit CKM matrix elements. Red is 90\% C.L. and cyan is
68\% C.L. Top-left panel corresponds to picking-up the parametrization 2 (eq. \ref{par5})
for both the u-type and d-type quarks. Top-right panel corresponds to parametrization 2 for the u-type quarks
and parametrization 3 (eq. \ref{par6}) for the d-type quarks.
Bottom-left panel corresponds to parametrization 3 for the u-type quarks and
parametrization 2 for d-type quarks. Finally, bottom-right panel shows the allowed region when
parametrization 3 is used for both u and d type quarks.}\label{chiCKM}
\end{figure}

Let us summarize the previous section. The real and symmetric matrix $\tilde H$ is diagonalized with the help of the eigenvectors
(\ref{eigenvectors}). The elements $a',b',c',d'$ are real and can be inverted with the help of the invariants (\ref{invariants}) into functions
of the masses $\tilde m_j$ and since we have only three invariants, we can choose $a'$ as a free parameter. Then,
$b'=b'(\tilde m_j,a'), c'=c'(\tilde m_j,a'), d'=d'(\tilde m_j,a')$ are functions of the masses and one free parameter $a'$. In order to have real elements,
$a'$ is restricted to a region delimited by the masses $\tilde m_j$. From all 32 possibilities of defining $b', c'$ and $ d'$ that
arises as solution of the set of equations given by the invariants, we found (\ref{par3},\ref{par5},\ref{par6}) as the only three
parametrizations that fulfill
all our requirements. That is: positive values $\tilde m_j$, eigenvalues of $\tilde H$ equal to $\tilde m_j^2$ and all elements of $\tilde H$ real.
The matrix $\mathcal{O}(\tilde m_j,a')$ that diagonalizes $\tilde H$ is defined by the eigenvectors (\ref{eigenvectors}).

The quark and lepton flavor mixing matrices, $U_{\rm PMNS}$ and $V_{\rm CKM}$ , arise from the mismatch between diagonalization
of the mass matrices of u and d type quarks and the diagonalization of the mass matrices of charged
leptons and left-handed neutrinos respectively.
Then, incorporating the phases again we have that the theoretical mixing matrix arising from four zero texture are:
\begin{equation}\label{theory_mixing}
V_{\rm CKM}^{\rm th} = O_u^T P^{u-d} O_d\,, \qquad  U_{\rm PMNS}^{\rm th} = O_l^T P^{l-\nu} O_\nu \,,
\end{equation}
where $P^{u-d}={\rm diag}[1,e^{i (\phi^u_1-\phi^d_1)},e^{i (\phi^u_2-\phi^d_2)}]$ and
in a similar way $P^{l-\nu}={\rm diag}[1,e^{i (\phi^\ell_1-\phi^\nu_1)},e^{i (\phi^\ell_2-\phi^\nu_2)}]$.


One can see that we have expressed $V_{\rm CKM}$ and $U_{\rm PMNS}$ as explicit
functions of the masses of quarks and leptons and few free parameters $a_u',a_d',
\phi,a_l', a_\nu',m_{\nu 3}$. For simplicity, and to restrict our space of parameters we
are going to fix $\phi\equiv\phi^u_1-\phi^d_1=\phi^u_2-\phi^d_2$ and $\Phi_1\equiv\phi^\ell_1-\phi^\nu_1=\phi^\ell_2-\phi^\nu_2$.
The masses of the quarks are well determined as well as the
masses of the charged leptons, namely (in MeV):
\begin{eqnarray}\label{quarkmasses}
2.35 < &m_u& < 4.14 \nonumber\\
730.5  < &m_c& < 789.5 \nonumber\\
1.59\times10^{5} <& m_t& < 1.83\times10^{5} \nonumber\\
3.76 <& m_d& < 5.04 \nonumber\\
94.0 <& m_s& < 106.0 \nonumber\\
2.81\times10^{3} <& m_b& < 3.03\times 10^{3}\,.
\end{eqnarray}
As we are interested by the mass contributions coming from Yukawa couplings, the masses (\ref{quarkmasses}) are calculated at the energy scale of the mass of the top quark
\cite{Fusaoka:1998vc,Leutwyler:1996qg,Pineda:1997hz} where the QCD interactions effects are well in the perturbative  regime. The running of the mixing matrices from experimental scale to top mass scale can be easily neglected \cite{Sasaki:1986jv,Babu:1987im,Luo:2009wa}.
The CKM matrix is one of the precision test of the standard model. The
precision in the determination of the parameters
have increased over the past decades. Current values are \cite{Nakamura:2010zzi}:
\begin{equation} \label{VCKM}
V_{\rm CKM}=
\begin{pmatrix}
0.97428 \pm 0.00015& 0.2253 \pm 0.0007 & 0.00347^{+0.00016}_{-0.00012}  \\
0.2252\pm 0.0007 & 0.97345^{+0.00015}_{-0.00016} & 0.0410^{+0.0011}_{-0.0007} \\
0.00862^{+0.00026}_{-0.00020} & 0.0403^{+0.0011}_{-0.0007} & 0.999152^{+0.000030}_{-0.000045}
\end{pmatrix}\,.
\end{equation}

\begin{figure}[t]
\includegraphics[width=0.7\textwidth]{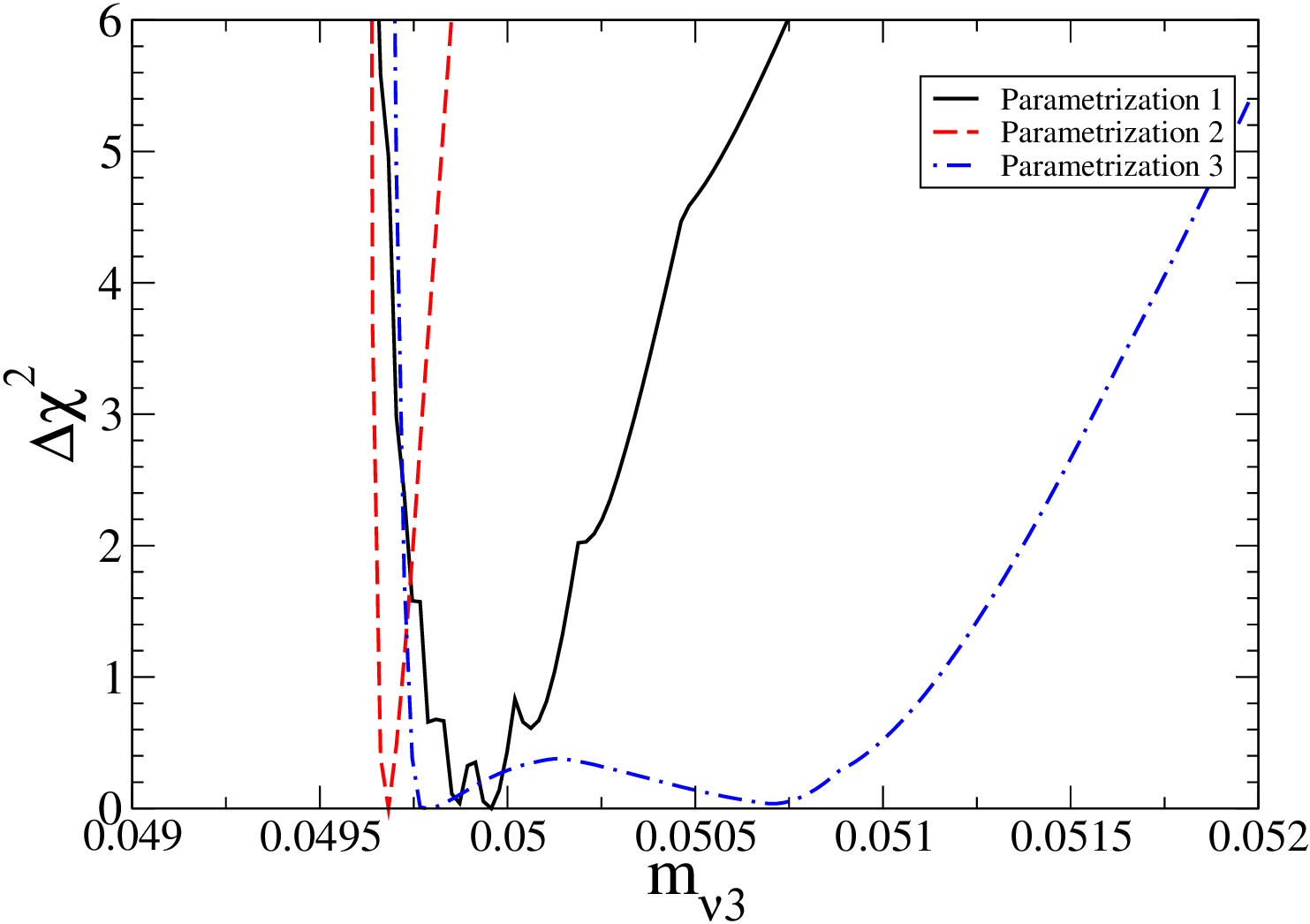}
\caption{Predicted mass for the third generation neutrino using the same parametrization
for both charged leptons and neutrinos.}\label{chi_masas}
\end{figure}

In addition to the moduli of the CKM matrix, we have information of the
angles $\alpha,\beta,\gamma$ which could include information of the phases
that can be lost if we consider only the moduli of the CKM elements since
\begin{eqnarray}
\alpha^{th}&=&{\rm arg}\left(-\frac{V_{cd}V_{cb}^*}{V_{td}V_{tb}^*}\right) \\
\beta^{th}&=&{\rm arg}\left(-\frac{V_{td}V_{tb}^*}{V_{ud}V_{ub}^*}\right) \\
\gamma^{th}&=&{\rm arg}\left(-\frac{V_{ud}V_{ub}^*}{V_{cd}V_{cb}^*}\right) \,.
\end{eqnarray}

The current values of the unitary angles are \cite{Nakamura:2010zzi}:
\begin{eqnarray}
\alpha^{Exp}&=& 89.0^o\pm  4.4^o \nonumber\\
\beta^{Exp}&=& 21.15 \pm 0.65^o \nonumber\\
\gamma^{Exp}&=& 73^o \pm 22^o
\end{eqnarray}

On the other hand, the neutrino masses are not measured. Instead,
the difference in masses have been obtained from the
observation of solar, atmospheric, reactor and accelerator neutrinos.
The current limits on the mass difference are
\begin{eqnarray}
\Delta m_{12}^2&=& 7.67^{+0.22}_{-0.21}\times 10^{-5} {\rm eV}^2\nonumber \\
\Delta m_{13}^2&=& 2.46 \pm 0.15 \times 10^{-3} {\rm eV}^2 \quad ({\rm normal \quad hierarchy})
\end{eqnarray}
and the determination of the mixing angles, including
the latest result for a large lepton mixing angle $\theta_{13}$ provided
by T2K, MINOS and Double Chooz experiments, gives updated values for
$U_{\rm PMNS}$ \cite{Fogli:2011qn}:

\begin{equation}\label{UPMNS}
U_{\rm PMNS}=
\begin{pmatrix}
0.824^{+0.011}_{-0.010}& 0.547^{+0.016}_{-0.014} & 0.145^{+0.022}_{-0.031}\\
0.500^{+0.027}_{-0.021} & 0.582^{+0.050}_{-0.023} & 0.641^{+0.061}_{-0.023} \\
0.267^{+0.044}_{-0.027} & 0.601^{+0.048}_{-0.022} & 0.754^{+0.052}_{-0.020} \,.
\end{pmatrix}
\end{equation}

With the the help of the definition of the theoretical mixing matrices
(\ref{theory_mixing}) and the experimental values (\ref{VCKM}) we perform
a simple $\chi^2$ analysis on the
\begin{equation}
\chi^2_{\rm quarks}(a_u,a_d,\phi)=\sum^9_{i=1}\left(\frac{V_{\rm CKM}^{th}(a_u,a_d,\phi)-|V_{\rm CKM}|}{\delta V_{\rm CKM}}\right)^2+\chi^2_{\rm angles}\,.
\end{equation}
with
\begin{equation}
\chi^2_{\rm angles}=\left(\frac{\alpha^{th}-\alpha^{Exp}}{\delta \alpha^{Exp}}\right)^2
+\left(\frac{\beta^{th}-\beta^{Exp}}{\delta \beta^{Exp}}\right)^2+\left(\frac{\gamma^{th}-\gamma^{Exp}}{\delta \gamma^{Exp}}\right)^2\,.
\end{equation}

For the case of $U_{\rm PMNS}$, the $\chi^2$ is also a function of the mass of the heaviest neutrino $m_{\nu 3}$.
For leptons we perform the $\chi^2$ analysis only with the moduli of the $U_{\rm PMNS}$ matrix
\begin{equation}
\chi^2_{\rm leptons}(a_l,a_\nu,\Phi_1, m_{\nu 3})=\sum^9_{i=1}\left(\frac{U_{\rm PMNS}^{th}(a_l,a_\nu,\Phi_1, m_{\nu 3})-|U_{\rm PMNS}|}{\delta U_{\rm PMNS}}\right)^2\,.
\end{equation}

\begin{figure}[t]
\includegraphics[width=0.8\textwidth]{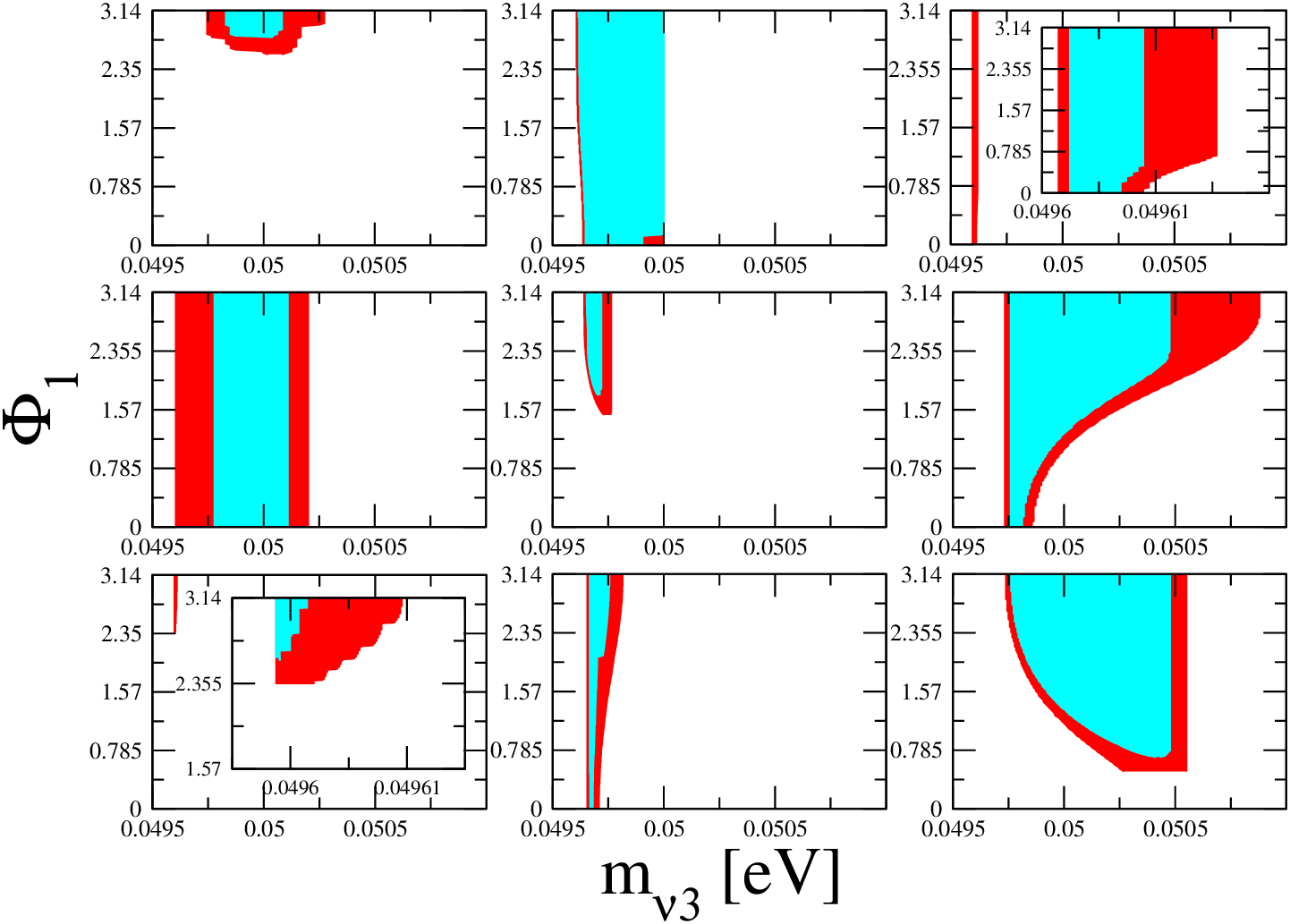}
\caption{$m_{\nu 3}-\Phi_1$ allowed regions that fit $U_{\rm PMNS}$ for different choices in the parametrization for charged
leptons and neutrinos. From left to right, and top to bottom: parametrization 1 for both, charged leptons and neutrinos,
parametrization 1 for charged leptons and parametrization 2 for neutrinos, parametrization 1 for charged leptons and parametrization 3
for neutrinos and so on.}\label{chi_Umns}
\end{figure}

The results are summarized in Figs. \ref{chiCKM}, \ref{chi_masas},\ref{chi_Umns} and Table \ref{tabla1} and Table \ref{tabla2}.

\begin{table}[!t]
\begin{tabular}{|c|c|c|c|}
\hline
Parametrization & 1 & 2 & 3 \\
\hline
1 &$\chi^2_{min}=930.02$ &$\chi^2_{min}=3.55\times10^{5}$ &$\chi^2_{min}=3.55\times10^5$ \\
\hline
2 &$\chi^2_{min}=1.43\times10^5$ &$\chi^2_{min}=1.270$ &$\chi^2_{min}=2.212$ \\
\hline
3 &$\chi^2_{min}=4.50\times 10^5$ &$\chi^2_{min}=1.44$ &$\chi^2_{min}=2.163$ \\
\hline
\end{tabular}
\caption{The minimum value of $\chi^2_{\rm quarks}$. We can see that some parametrizations
produces very big values $\chi^2_{\rm quarks}$, and so, they are ruled out as good parametrization}\label{tabla1}
\end{table}

Fig. \ref{chiCKM} shows the allowed values of $a_u'$ and $a_d'$ that reproduce the $V_{\rm CKM}$ matrix elements
for different parametrization for $u$ and $d$-type quarks.
Additionally, we can see from the minimum values of $\chi^2_{\rm quarks}$ reported in Table \ref{tabla1}
that some parametrization are not able to allow us to reproduce $V_{\rm CKM}$ or the angles. That is the reason
why only four different combinations are shown in Fig. \ref{chiCKM}.

On the other hand, for the case of neutrinos, the minimum $\chi^2_{\rm leptons}$ is reported in Table \ref{tabla2}.
It shows that the nine different combinations of parametrization for charged leptons and neutrinos give
reasonable values of $\chi^2_{\rm leptons}$. Fig. \ref{chi_masas} shows the projection in $m_{nu_3}$ of the neutrino
mass for the case when the same parametrization for both charged leptons and neutrinos.
Finally, Fig. \ref{chi_Umns} shows the projection in the plane $m_{\nu 3}-\Phi_1$ at 68\% C.L. (cyan)
and 90\% C.L. (red).

\begin{table}[!t]
\begin{tabular}{|c|c|c|c|}
\hline
Parametrization & 1 & 2 & 3 \\
\hline
1 &$\chi^2_{min}=4.758$ &$\chi^2_{min}=0.009$ &$\chi^2_{min}=11.02$ \\
\hline
2 &$\chi^2_{min}=4.798$ &$\chi^2_{min}=0.006$ &$\chi^2_{min}=3.119$ \\
\hline
3 &$\chi^2_{min}=7.087$ &$\chi^2_{min}=0.456$ &$\chi^2_{min}=0.006$ \\
\hline
\end{tabular}
\caption{The minimum value of $\chi^2_{leptons}$.
produces very big values $\chi^2_{quarks}$, contrary to the case of quarks, most of the parametrization
give reasonable values of $\chi^2_{min}$}\label{tabla2}
\vspace{.5cm}
\end{table}

\section{Conclusion}\label{conclusions}
To conclude, we want to stress that this is a  complete analysis for the four-zero textures
in quark and leptonic sectors for no-hermitian Yukawa matrices made within a general formalism.In order to take into account the CP violation, one assumption has been done in order to simplify the inclusion of the CP violating phases in  our analysis. This assumption  corresponds  to fix $\phi\equiv\phi^u_1-\phi^d_1=\phi^u_2-\phi^d_2$ and $\Phi_1\equiv\phi^\ell_1-\phi^\nu_1=\phi^\ell_2-\phi^\nu_2$ as defined in eq.(15).  The diagonalization of the mass matrices has been obtained without introducing any more approximations.
We analyse   the different  parametrizations of the four-zero texture mass matrices for quarks and leptons,
that are able to reproduce the CKM and PMNS mixing matrices. It is important to stress that the relation between fermion masses and mixing angles obtained through our analysis are not the same as the ones usually given in Hermitian case  where some  approximation have been done in order to diagonalize the mass matrices \cite{Matsuda:2006xa} \footnote{ Of course, if one assumes that our Yukawa matrices are hermitian, one can recover the usual relation between fermion masses and mixing angles}. This analysis is done through a $\chi^2$ analysis
with up to date values of the mixing matrices and angles.
In quark sector, only four solutions are found to be compatible with CKM mixing matrix. In leptonic sector,
using the last experimental results about the mixing angles in the neutrino sector, our $ \chi^2$ analysis shows
a preferred value for $m_{\nu_3}$ to be around $0.05$ $eV$ independently of the parametrization of the four-zero
texture mass matrices chosen for the charged leptons and neutrinos. This is a strong prediction for the
four-zero texture models. This value for neutrino masses favors standard  leptogenesis as the mechanism
to produce the Baryon Asymmetry of the Universe \cite{Buchmuller:2005eh}.
As expected, the leptonic $CP$ violating phases cannot be fixed through this analysis.

\acknowledgments
D.D. is grateful to DAIP project (Guanajuato University) and to CONACYT project for their financial support.
This work has been partially supported by Conacyt SNI-Mexico and PIFI funds (SEP, Mexico). We thank F. Gonzalez-Canales for helpful discussions.
L.L.L. thanks CONACYT for financial support.
\bibliography{bibliography}
\end{document}